\documentclass[twocolumn,aps,prl,amssymb,superscriptaddress, showkeys,floatfix]{revtex4-1}
\usepackage{amsmath}
\usepackage{graphicx}
\DeclareGraphicsExtensions{.pdf,.eps,.png,.jpg,.mps}
\usepackage{epstopdf}
\usepackage{threeparttable}
\usepackage{multirow}
\usepackage{xcolor}
\usepackage{url}
\usepackage{afterpage}
\begin{document}
\title{Graphene Overcoats for Ultra-High Storage Density Magnetic Media}
\author{N. Dwivedi$^1$*, A. K. Ott$^2$*, K. Sasikumar$^3$, C. Dou$^2$, R. J. Yeo$^{1,4}$, B. Narayanan$^3$, U. Sassi$^2$, D. De Fazio$^2$, G. Soavi$^2$, T. Dutta$^{1,5}$, S. K. R. S. Sankaranarayanan$^3$, A. C. Ferrari$^2$, C. S. Bhatia}
\affiliation{Department of Electrical and Computer Engineering, National University of Singapore, Singapore 117583\\
$^2$Cambridge Graphene Centre, University of Cambridge, Cambridge CB3 0FA, UK\\
$^3$Center for Nanoscale Materials, Argonne National Laboratory, 9700 S Cass Avenue, Argonne IL USA\\
$^4$Institute of Materials, Ecole Polytechnique F\'ed\'erale de Lausanne (EPFL), 1015 Lausanne, Switzerland\\
$^5$ Institute of Materials Research and Engineering, A*STAR (Agency for Science, Technology and Research), 3 Research Link, Singapore 117602, Singapore\\
* These authors contributed equally}
\begin{abstract}
Hard disk drives (HDDs) are used as secondary storage in a number of digital electronic devices owing to low cost ($<$0.1\$/GB at 2016 prices) and large data storage capacity (10TB with a 3.5 inch HDD). Due to the exponentially increasing amount of data, there is a need to increase areal storage densities beyond$\sim$1Tb/in$^2$. This requires the thickness of carbon overcoats (COCs) to be$<$2nm. Friction, wear, corrosion, and thermal stability are critical concerns$<$2nm, where most of the protective properties of current COCs are lost. This limits current technology and restricts COC integration with heat assisted magnetic recording technology (HAMR), since this also requires laser irradiation stability. Here we show that graphene-based overcoats can overcome all these limitations. 2-4 layers of graphene enable two-fold reduction in friction and provide better corrosion and wear than state-of-the-art COCs. A single graphene layer is enough to reduce corrosion$\sim$2.5 times. We also show that graphene can withstand HAMR conditions. Thus, graphene-based overcoats can enable ultrahigh areal density HDDs$>$10Tb/in$^2$.
\end{abstract}
\maketitle
\section{\label{In}Introduction}
There has been an incessant increase in data generation over the past few decades, owing to enhanced use of personal computers\cite{Reinsel2017}, digital electronic products\cite{Reinsel2017}, and the continuous growth of automation\cite{Reinsel2017}. The annual data creation rate was$\sim$16.1zettabytes/year (ZB, 1ZB$=$trillion gigabytes (GB)) in 2016\cite{Reinsel2017}, and is expected to increase to$\sim$163ZB/year by 2025\cite{Reinsel2017}. While various devices are used to store digital information, such as tape\cite{tape} and flash drives\cite{Grochowski,HDDmarket}, hard disk drives (HDDs) remain the primary choice as secondary storage device, i.e. as external storage device with a non-volatile memory function that holds data until it is deleted or overwritten, in contrast to primary storage devices, e.g. random access memory or cache, that hold data only temporarily while the computer is running. This is due to their low cost($<$0.1\$/GB at 2016 prices\cite{Grochowski}) and large storage capacity$>$10TB with 3.5 inch HDDs\cite{Grochowski}. This is reflected in the 425 million HDDs shipped globally in 2016\cite{HDDmarket}, decreasing to 350 million by 2020\cite{HDDmarket}, due to the expected increase in storage capacity per drive. HDDs will rule storage technologies at least for the next 5-10 years in terms of capacity\cite{Grochowski,HDDmarket}, price\cite{Grochowski,HDDmarket}, production\cite{Grochowski,HDDmarket} and shipment\cite{Grochowski,HDDmarket}. Desktops, laptops, cloud computing, consumer digital electronic products predominantly employ HDDs\cite{Piramanayagam,McFadyen,Mao}. A HDD contains two major components: 1) hard disk medium (HDM)\cite{Piramanayagam,McFadyen}; 2) head\cite{McFadyen,Mao}. In a HDD the information is written in/read from the HDM by the head\cite{Zhang2010,Plumer2011}. To meet the rise of data storage demand, HDD technology has moved from inductive, used prior to 1990\cite{McFadyen,Bajorek2014}, to magneto-resistive from 1990\cite{McFadyen,Bajorek2014}, and giant MR (GMR) or tunneling MR (TMR) heads from 1997\cite{McFadyen,Mao,Bajorek2014} and 2004 onwards\cite{McFadyen,Mao,Bajorek2014}. HDM technology moved from particulate media\cite{Piramanayagam} in the first generations HDDs in the late 1950s\cite{Piramanayagam} to thin film media\cite{Kryder1992, Piramanayagam} after their introduction in the 1980s into 5.35inch drives\cite{Kryder1992, Piramanayagam}. Servo and signal processing technologies have also advanced\cite{Zhang2010}. Solid state drives (SSDs)\cite{Grochowski,HDDmarket} are the main competing technology, posing a threat to future HDD viability\cite{HDDmarket}. The main advantage of HDDs is that they can store a large amount of data$\sim$TB cheaply. As of 2019, SSDs are still more than two times more expensive than HDDs: A 1TB internal 2.5-inch hard drive costs$\sim$\$40-50\cite{SSD-HDD}, while an SSD of the same capacity and form factor starts at$\sim$\$250\cite{SSD,SSD-HDD}. The main advantage of SSDs over HDDs is that they do not have mechanically moving parts, making them more stable. The second advantage is that SSDs based on NAND flash memories are faster:$\sim$10-13s average bootup time for SSDs\cite{SSD-HDD}, compared to$\sim$30-40s for HDDs\cite{SSD-HDD}. Thus, novel technologies are needed to enable high areal density (AD) HDDs with$>$1Tb/in$^2$\cite{Grochowski,10Marchon2013}.

One option is to reduce the head-media spacing (HMS), since this reduces the signal-to-noise ratio\cite{10Marchon2013,Wallace2000} and limits the AD growth\cite{10Marchon2013,Wallace2000}. There are many contributors to the HMS, and for 1Tb/in$^2$, the total HMS is$\sim$8.9-6.5nm\cite{10Marchon2013}, based on the sum of HDM overcoat thickness (2.5$-$2.0nm)\cite{10Marchon2013}, lubricant thickness (1.2$-$1.0nm)\cite{10Marchon2013}, touch down height (2.0$-$1.0nm)\cite{10Marchon2013}, fly clearance (1.2$-$1.0nm)\cite{10Marchon2013} and head overcoat thickness (2.0$-$1.5nm)\cite{10Marchon2013}. Thus, the HDM overcoat is the largest contributor. This comprises magnetic/metallic layers, including Co-alloys-based magnetic storage layers\cite{Piramanayagam,Casiraghi2007}, with a high coefficient of friction COF, i.e. ratio of frictional force to normal force\cite{friction-book},$\sim$0.6-0.8\cite{Dwivedi2015,15Dwivedi2015b} and wear\cite{Dwivedi2015,15Dwivedi2015b}. Therefore, they can experience mechanical damage whenever intermittent contact occurs with the head\cite{Dwivedi2015,15Dwivedi2015b}, and are susceptible to corrosion\cite{Casiraghi2007,Dwivedi2015,Casiraghi2004b} and oxidation\cite{Dwivedi2015,15Dwivedi2015b}, leading to HDD degradation or damage. Carbon-based overcoats (COCs) are widely used to protect HDM from mechanical damage\cite{Casiraghi2004a,Casiraghi2004b, Piazza2004,Piazza2005, Dwivedi2015,Rajauria2015} and corrosion\cite{Casiraghi2004b,Ferrari2004, Casiraghi2007,15Dwivedi2015b}, to ensure maintained functionality and durability\cite{Casiraghi2007, Ferrari2004}. The COC thickness reduced from 12.5nm in 1990\cite{Yogi1990}, corresponding to 1Gb/in$^2$\cite{Yogi1990}, to 2.5$-$3nm in 2013 for 0.8$-$1Tb/in$^2$\cite{10Marchon2013,Dwivedi2015}, with a planned reduction to 1.8-1.5nm for 4Tb/in$^2$ by 2020-2021\cite{10Marchon2013}. By extrapolating Fig.2 in Ref.\cite{10Marchon2013},$<$1nm would be required for$\sim$10Tb/in$^2$. However, current COCs lose most of their appealing properties, such as anti-friction\cite{Dwivedi2015,15Dwivedi2015b,Rajauria2015}, wear resistance\cite{Dwivedi2015,Rajauria2015}, Young's Modulus\cite{Casiraghi2004a,Casiraghi2004b,Casiraghi2005,Ferrari2004}, and corrosion protection\cite{Casiraghi2004a,Casiraghi2004b, Casiraghi2005,Ferrari2004},$<$2-3nm. The current$\sim$2.7nm commercial COCs have high COF$\sim$0.3-0.5\cite{Dwivedi2015,15Dwivedi2015b} and wear in a ball-on-disk tribological environment\cite{Dwivedi2015,15Dwivedi2015b}, which can result in damage, hence durability concerns. The ideal overcoat needs to provide: 1) corrosion protection, which requires complete coverage\cite{Casiraghi2007,Dwivedi2015} and a layer without any pinholes\cite{Ferrari2004,Casiraghi2007}; 2) COF$<$0.3-0.5, lower than in 2.7nm commercial COCs\cite{Dwivedi2015,15Dwivedi2015b} and wear resistance\cite{Casiraghi2007,Dwivedi2015,Rajauria2015}, which requires better lubricating properties\cite{Rajauria2015,Dwivedi2015}, hardness and elasticity with a Young's modulus of at least$\sim$400GPa\cite{Casiraghi2007,Casiraghi2004a}; 3) lubricant compatibility\cite{Casiraghi2007}; 4) surface smoothness\cite{Moseler2005,Casiraghi2005,Pisana2006}, e.g. root mean square roughness$\sim$0.2-0.3nm\cite{Casiraghi2007}. Given the limitations to achieve most of these properties with conventional COCs$<$2nm\cite{Casiraghi2004a, Casiraghi2004b,Ferrari2004}, either the search of novel overcoats or engineering of existing overcoats is required to enable future ultra-high AD HDDs.

Another critical bottleneck hindering AD growth is the superparamagnetic limit\cite{Piramanayagam,Plumer2011,Richter2007}. In order to increase AD, the grain size of the magnetic recording layer needs to be$<$8-9nm\cite{Plumer2011, Piramanayagam}. Between 2000 and 2010 this was reduced from 20 to 9nm\cite{Plumer2011}, to enable denser packing of information\cite{Piramanayagam,Plumer2011,Richter2007}. However, depending on material, superparamagnetism comes into play for grain sizes$<$8-9nm\cite{Plumer2011,Richter2007,Plumer2001}. The reduction of grain size results in a decrease of their magnetic anisotropy energy\cite{Plumer2011,Richter2007,Weller1999}, and the magnetization can randomly flip in response of thermal fluctuations\cite{Plumer2011,Richter2007}. For thermal stability of small grains$\sim$3nm over long periods of time (up to 10 years) the magnetic recording layer must have K$_{\text{u}}$V/k$_{\text{B}}$T$\geq40-60$\cite{Plumer2011,Plumer2001,Weller1999}, where K$_u$=1/2H$_K$M\cite{Plumer2011} is the anisotropy constant, H$_K$ [A/m] the magnetic anisotropy field, M [A/m] the magnetization, V [m$^3$] is the grain volume, k$_B$ the Boltzmann constant and T the temperature. Thus, magnetic materials with large$>$100kOe H$_K$ are needed to support grains down to 3nm with stable magnetization, i.e. not in the superparamagnetic limit. E.g., FePt supports stable grains with diameters down to 2-3nm\cite{Plumer2001,Weller1999,Kryder2008,Challener2009}. However, higher H$_K$ results in higher coercivity, i.e. the resistance of a magnetic material to changes in the magnetization, which is equivalent to the field intensity necessary to demagnetize the fully magnetized material, e.g. up to 50000Oe for FePt\cite{Plumer2001} (10000Oe=1T), much larger than in today's recording heads with$\sim$1-2T\cite{Plumer2011,Plumer2001}. This introduces issues for FePt-based HDM, because the writing field must be comparable to the coercivity of the magnetic storage layer in order to encode information\cite{Plumer2011}. These concerns are usually called the "magnetic trilemma"\cite{Plumer2011}. To overcome these issues, heat assisted magnetic recording (HAMR) was suggested\cite{Weller1999,Kryder2008,Challener2009}. HAMR uses a laser to heat the magnetic medium for$\sim$1ns\cite{Weller1999} in order to decrease its coercivity, bringing the material to its Curie temperature T$_C$, i.e. the T above which the material loses permanent magnetic properties. Above T$_C$, the HDM coercivity becomes comparable to/lower than the head field\cite{Weller1999,Kryder2008,Challener2009,Hu2017,Vogler2016}, where it is writable with a magnetic field$\sim$0.8T by conventional heads\cite{Vogler2016}. After cooling to room T, the coercivity goes back to its original value, and data are retained\cite{Plumer2001}.

While HAMR appears to be a solution to all issues, such as small grain size, high magnetic anisotropy and writability, and paves the way to AD increase, it raises concerns on the thermal stability of COCs against laser irradiation\cite{Jones2014,Ji2014,Mangolini2013}. As T$_C\sim$700-750K for FePt\cite{Jones2014,Ji2014,Weller2016} is reached by laser irradiation, the overcoats for HAMR require thermal stability at least up to$\sim$700-800K to avoid degradation over time\cite{Weller1999,Jones2014,Ji2014}. Efforts are ongoing to test the thermal stability of HDM overcoats in variable environments, e.g. by using heating techniques such as laser annealing\cite{Jones2014, Ji2014}, thermal annealing\cite{Mangolini2013} and by changing the thermal treatment time\cite{Jones2014, Mangolini2013, Wang2013} and heating rate\cite{Jones2014, Ji2014, Mangolini2013, Wang2013}, so as to design HAMR-compatible overcoats. Ref.\cite{Jones2014} reported oxidation, degradation, and removal of a$\sim$5nm hydrogenated amorphous carbon, a-C:H,-based commercial COC (on HAMR-compatible HDM) in HAMR-like conditions, where a pump laser (527nm) was used to heat the sample at a rate$\sim$10$^6$K/s\cite{Jones2014}, and the exposed area ($\sim$8$\mu$m) was analyzed with a probe laser (CW at 476nm, pulsed with chopper) for Raman measurements with 0.25s total irradiation time. Ref.\citenum{Ji2014} also reported degradation of a 4nm thick a-C:H-based commercial COC (on FePt-based HDM) in HAMR-like conditions, when using a 785nm laser, with a spot size$\sim$1$\mu$m, power$\sim$57mW for 1.5ns, with an additional 2.5ns to reach the peak power and another 2.5ns to go back to 0mW, and a cooling time$\sim$21.8ns before the next heating cycle. The total heating time in Ref.\cite{Ji2014} was$\sim$0.1ms, corresponding to a 5 year drive life or 157.68$\times$10$^6$s. For an HDD with 0.5$\times$10$^6$ tracks, 100 rotation/s and laser heating time 1.5ns, the total irradiation time for 5 years is 47.3$\times$10$^3$ns=47.3$\times$10$^{-3}$ms, i.e. less than 0.1ms. In general, commercial a-C:H COCs have shown degradation or structural changes in various heating environments such as laser irradiation\cite{Jones2014,Ji2014}, annealing at 773K at a rate$\sim$5K/s\cite{Mangolini2013}, rapid thermal annealing$\sim$380K/s to$\sim$930K for$<$1s\cite{Wang2013}. Better thermal stability was reported in filtered cathodic vacuum arc (FCVA)-based COCs under laser irradiation in HAMR-like\cite{Ji2014,Kundu2015} and thermal annealing (up to$\sim$940K) conditions\cite{Pathem2014}, for thickness$<$5nm, consistent with the good thermal stability, i.e. no change in sp$^3$ content up to 1100$^{\circ}$C, found in$\sim$70nm ta-C films\cite{Ferrari1999}. However, FCVA-based COCs are not yet used due to the presence of macro-particles\cite{10Marchon2013,Teo2002}. Overall, commercial COCs are facing several critical concerns which constitute a barrier to advance existing technology, and may not be integrated with future HAMR technology.

Graphene is an emerging material for lubrication\cite{Berman2015,Berman2014,Berman2014b,Berman2014c,Lee2010,Egbert2014}, as well as oxidation\cite{Topsakal2012,Chen2011a} and corrosion protection\cite{Martin2015,Weatherup2015,Prasai2012,Raman2012,Kravets2014}. Ref.\citenum{Berman2014b} reported that single layer graphene (SLG) reduced steel COF from 0.9 to 0.3 with a coating lifetime up to 6500 cycles and decreased wear rate by two orders of magnitude. Multilayer graphene (MLG) (3-4 layers) showed excellent tribological performance with COF$<$0.2, decrease of wear rate by three orders of magnitude and long sliding lifetime up to 47000 cycles on steel\cite{Berman2014b}. Ref.\cite{Berman2015} indicated that SLG exhibits superlubricity, i.e. friction nearly vanishes when used with nanodiamond on Si/SiO$_2$\cite{Berman2015}, and reduced friction by two to three times and wear rate by two orders of magnitude for Au-based electrical contacts\cite{Berman2014c}. SLG was reported to decrease the oxidation and corrosion of Ni\cite{Chen2011a,Prasai2012,Martin2015,Weatherup2015}, Co\cite{Weatherup2015}, Fe\cite{Weatherup2015}, Pt\cite{Weatherup2015}, Cu\cite{Prasai2012,Raman2012,Kravets2014,Chen2011a} and Ag\cite{Kravets2014}. Reduced graphene oxide (RGO) was also used as a barrier coating\cite{Su2014}. Refs.\cite{Yu2015,Kim2010} reported that SLG has good thermal stability up to 2600K, with thermal conductivity up to$\sim$2000W/mK\cite{Yu2015,Lee2011, Balandin2008, Ghosh2008,Faugeras2010,Seol2010,Cai2010,Chen2011}. All these characteristics make SLG promising as protective overcoat for both existing and HAMR-based technologies. Current commercial HDM use a layer of perfluoropolyether (PFPE) lubricant (lube)\cite{Dwivedi2015,Waltman2012} to further reduce friction and wear and minimize surface energy. Given the lubricating and corrosion protection properties of SLG, the need of lube for SLG is unclear. However, exploring compatibility of SLG with lube is critical from the fundamental viewpoint.

Here we use 1-4 layers of chemical vapor deposition (CVD) grown graphene transferred on Co-alloy (current technology) and FePt-based (HAMR technology) HDMs, and test friction, wear, corrosion and thermal stability, as well as lube compatibility. We demonstrate that SLG shows better performances than current 2.7nm COCs on Co-alloy-based HDM, and good tribological properties with a COF$<$0.2 for$>$10000 cycles. Thermal stability tests confirm that SLG on FePt HDM can withstand HAMR-like conditions, without degradation. SLG's superior performance and its thinness can enable the development of ultrahigh density magnetic data storage, based either on current technology, or on HAMR, or on HAMR combined with bit patterned media (BPM), where the storage layer is patterned into an array of pillars, each representing a single bit\cite{bit-patterned}. SLG+HAMR+BPM is expected to increase AD$>$10Tb/in$^2$.
\begin{figure*}	
\centerline{\includegraphics[width=160mm]{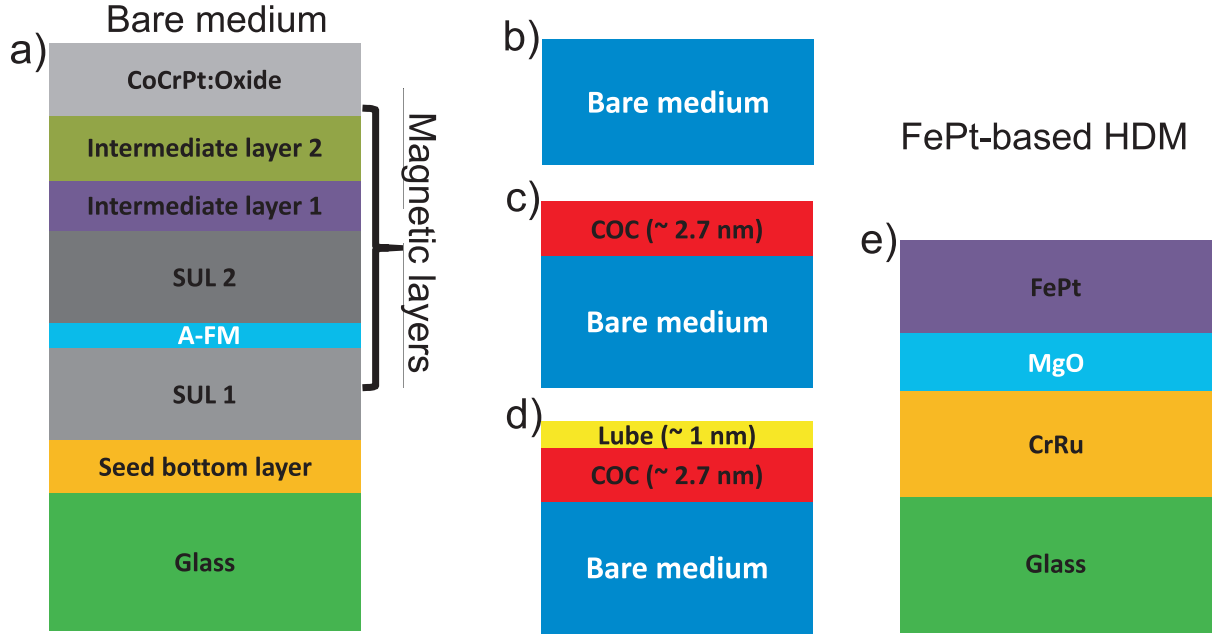}}
\caption{a) Full stack of bare CoCrPt:Oxide-based HDM with soft-magnetic underlayers (SUL) and an antiferromagnetic layer (A-FM). b) bare medium. c) magnetic medium with$\sim$2.7nm commercial COC. d) Magnetic medium with$\sim$2.7nm commercial COC and$\sim$1nm commercial lube. e) Full stack of bare FePt-based HDM}
\label{fig:fig1}
\end{figure*}
\begin{table*}[ht]
\caption{Samples summary}
\begin{tabular}{|l|l|}\hline
BM           & CoCrPt-oxide based BM without COC and lube\\\hline
BML          & CoCrPt-oxide based BM without COC, with PFPE lube (1.2$\pm$0.2nm)\\\hline
CMC          & CoCrPt-oxide based commercial magnetic media with commercial COC ($\sim$2.7nm) without lube\\\hline
CMCL         & CoCrPt-oxide based commercial magnetic media with commercial COC ($\sim$2.7nm) and commercial PFPE lube ($\sim$1nm)\\\hline
1LG         & SLG on CoCrPt-oxide based commercial magnetic media\\\hline
2LG         & 2LG on CoCrPt-oxide based commercial magnetic media\\\hline
3LG         & 3LG on CoCrPt-oxide based commercial magnetic media\\\hline
4LG         & 4LG on CoCrPt-oxide based commercial magnetic media\\\hline
1LGL        & PFPE lube (1.2$\pm$0.2)nm coated SLG on CoCrPt-oxide based commercial magnetic media\\\hline
2LGL        & PFPE lube (1.2$\pm$0.2)nm coated 2LG on CoCrPt-oxide based commercial magnetic media\\\hline
3LGL        & PFPE lube (1.2$\pm$0.2)nm coated 3LG on CoCrPt-oxide based commercial magnetic media\\\hline
4LGL        & PFPE lube (1.2$\pm$0.2)nm coated 4LG on CoCrPt-oxide based commercial magnetic media\\\hline
3CF         & FCVA-deposited ta-C ($\sim$0.3nm) on CoCrPt-oxide based commercial magnetic media\\\hline
6CF         & FCVA-deposited ta-C ($\sim$0.6nm) on CoCrPt-oxide based commercial magnetic media\\\hline
12CF        & FCVA-deposited ta-C ($\sim$1.2nm) on CoCrPt-oxide based commercial magnetic media\\\hline
18CF        & FCVA-deposited ta-C  ($\sim$1.8nm) on CoCrPt-oxide based commercial magnetic media\\\hline
8CS         & Pulsed DC sputtered sp$^2$ rich carbon ($\sim$0.8nm) on CoCrPt-oxide based commercial magnetic media\\\hline
12CS        & Pulsed DC sputtered sp$^2$ rich carbon ($\sim$1.2nm) on CoCrPt-oxide based commercial magnetic media\\\hline
\end{tabular}
\label{tab:samples}
\end{table*}
\begin{figure*}
\centerline{\includegraphics[width=170mm]{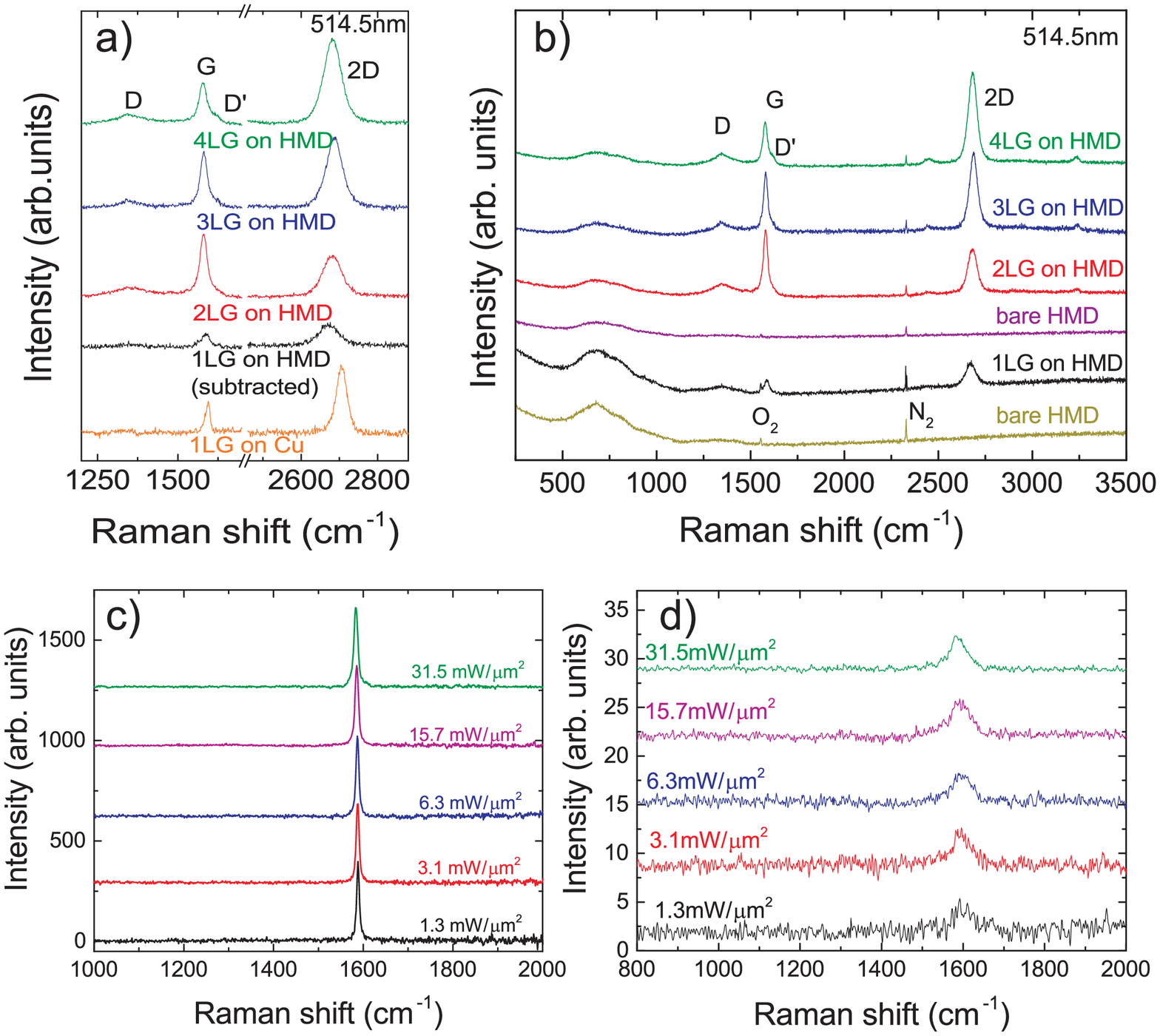}}
\caption{(a) Raman spectra of as-grown graphene on Cu (before transfer) and transferred on CoCrPt:Oxide after CoCrPt:Oxide background subtraction. (b) Raman spectra of CoCrPt:Oxide and 1LG to 4LG transferred onto it. All Raman spectra in (a) and (b) are measured at 514.5nm. Raman spectra of 1LG: (c) on Si/SiO$_2$ and (d) on FePt-based (HAMR) HDM for different laser power densities at 785nm excitation}
\label{fig:fig2}
\end{figure*}
\section{Results}
\subsection{Magnetic Media Substrates}
We use CoCrPt:Oxide-based bare HDMs (BM) from Hitachi\cite{Waltman2012}. These comprise multiple layers: seed bottom layer, soft-magnetic under layers (SUL), antiferromagnetic layer (A-FM), intermediate layers and CoCrPt:Oxide storage layer, Figs.\ref{fig:fig1}a,b. We also use other HDMs from HGST\cite{Waltman2012}: HDM with$\sim$2.7nm commercial COC (CMC, Fig.\ref{fig:fig1}c), and$\sim$2.7nm commercial COC+$\sim$1nm commercial PFPE lube (CMCL, Fig.\ref{fig:fig1}d). These samples are used as reference. The commercial COC is plasma-assisted CVD (PACVD) grown a-C:H followed by a nitrogen plasma surface treatment\cite{Dwivedi2015,Goohpattader2015,Jones2014,Ji2014,Mangolini2013,Wang2013,Waltman2012}. PACVD, magnetron sputtering and FCVA are commonly used to deposit COCs\cite{Casiraghi2007,Dwivedi2015,15Dwivedi2015b,Rajauria2015,Ferrari2004,Casiraghi2004b,Dwivedi2016c,Goohpattader2015,Jones2014,Ji2014,Mangolini2013,Wang2013,Kundu2015,Pathem2014,Waltman2012}. We also use magnetron sputtering\cite{Kundu2015} and FCVA to deposit and compare the performance of all types of COCs with that of 1-4LG. COCs with thicknesses from$\sim$0.3 to$\sim$1.8nm are deposited on Co-alloy-based HDM using pulsed DC magnetron sputtering\cite{Dwivedi2016c,Dwivedi2015c} and FCVA\cite{Casiraghi2007,Dwivedi2015,Ferrari2004,Dwivedi2016c,Goohpattader2015,Kundu2015,Pathem2014,Ferrari1999,Polo2000,Fallon1993}. A first COC set is deposited using the NTI Media coating system FS2 FCVA, equipped with a double bend macro-particles filter\cite{Goohpattader2015} at an ion energy and dose of 50eV and 3.75$\times$10$^{15}$ ions/cm$^2$. The thickness is controlled by varying the deposition time and the rate is calibrated using atomic force microscopy (AFM) and high resolution transmission electron microscopy (HRTEM)\cite{Goohpattader2015}. A second set of COCs is prepared using an AJA sputtering tool\cite{Dwivedi2016c,Dwivedi2015c}. Argon (Ar) plasma cleaning is performed for 3mins, prior to COC deposition, at RF power$\sim$40W, Ar gas flow rate$\sim$20sccm and pressure$\sim$10mTorr to remove surface contamination\cite{Dwivedi2015c}. 99.999\% pure graphite is used as target. The COCs are grown in an Ar atmosphere with pressure and gas flow rate$\sim$5mTorr and 20sccm. A pulsed DC power$\sim$100W with 150kHz frequency and duration$\sim$2.6$\mu$s is supplied to the graphite target. The deposition time is varied to achieve different thicknesses. The details and nomenclature of the different COCs are listed in Table \ref{tab:samples}. The structural properties of these FCVA and sputter deposited COCs can be found in Refs.\cite{Goohpattader2015,Dwivedi2015c}. HAMR-compatible FePt-based HDMs are prepared at a base pressure$\sim$1.8$\times$10$^{-9}$Torr by magnetron sputtering. These are stacks of FePt/MgO/CrRu/glass, Fig.\ref{fig:fig1}e, where CrRu and MgO seed layers\cite{Weller2016} are grown at$\sim$400$^{\circ}$C and$\sim$1.5mTorr. The FePt storage layer is grown at$\sim$600$^{\circ}$C and 3.5mTorr, Table \ref{tab:samples}.
\subsection{Graphene growth and transfer}
SLG is grown by chemical vapor deposition (CVD) on a 35$\mu$m Cu foil following a similar procedure to Ref.\cite{Bae2010}. The substrate is loaded into a hot wall tube furnace at$\sim$1mTorr. The Cu foil is annealed in hydrogen at 1000$^{\circ}$C for 30mins. This reduces the oxide surface and increases the Cu grain size\cite{Bae2010}. The growth process starts when 5sccm CH$_4$ is added. After 30mins the substrate is cooled-down in vacuum ($\sim$1mTorr) and unloaded.

The quality and uniformity of the samples is assessed by Raman spectroscopy. Unpolarized spectra are recorded at 514.5 as well as at 785nm, close to 850nm, used for HAMR writing tests, with a Renishaw InVia spectrometer equipped with a Leica DM LM microscope and a 100x objective with a numerical aperture 0.85. A 514.5nm Raman spectrum of 1LG on Cu before transfer is shown in Fig.\ref{fig:fig2}a. The photoluminescence (PL) background due to the Cu foil is removed using baseline subtraction\cite{Lagatsy2013}. The D to G intensity ratio I(D)/I(G)$<$0.1 indicates a defect concentration n$_{\text{d}}<2.4\times10^{10}$cm$^{-2}$\cite{Ferrari2013,Cancado2011,Ferrari2000,Bruna2014}, Fig.\ref{fig:fig2}b. The 2D peak position can be fitted with a single Lorentzian with Pos(2D)=2705cm$^{-1}$ and full-width-half-maximum, FWHM(2D)$\sim$33cm$^{-1}$, a signature of SLG\cite{Ferrari2006, Ferrari2013}. The G peak position and FWHM(G) are$\sim$1593cm$^{-1}$ and 16cm$^{-1}$. I(2D)/I(G), A(2D)/A(G) are$\sim$2.3 and 4.9.

After Raman characterization, 1LG is placed onto CoCrPt:Oxide. We use a polymethyl methacrylate (PMMA)-based wet transfer\cite{Bonaccorso2012,Bonaccorso2010}. First,$\sim$500nm PMMA is spin coated on the sample. The PMMA/SLG/Cu stack is then placed in an aqueous solution of ammonium persulfate to etch Cu\cite{Bonaccorso2012}. When Cu is fully etched, the graphene/PMMA-stack is placed into de-ionized (DI) water to rinse any acid residuals, and subsequently fished out using the HDM. After drying for one day at room T, the sample is placed in acetone to remove PMMA, leaving 1LG on the HDM. By repeating the steps described above, several 1LGs are transferred to create a MLG stack. The same procedure is used to place 1LG and MLG onto FePt.

Representative Raman spectra of 1-4LG on HDM are in Fig.\ref{fig:fig2}a, with HDM spectrum subtracted. This and the spectra of 1L-4LG on HDM before background subtraction are in Fig.\ref{fig:fig2}b. By subtracting the reference spectrum, using the N$_2$ Raman peak from air for normalization, from the spectrum taken on 1LG-coated HDM, we can reveal the 1LG contribution. After transfer, all peaks are downshifted and broader, with Pos(2D)$_{\text{1LG}}\sim$2672cm$^{-1}$, Pos(2D)$_{\text{2LG}}\sim$2681cm$^{-1}$, Pos(2D)$_{\text{3LG}}\sim$2687cm$^{-1}$, Pos(2D)$_{\text{4LG}}\sim$2682cm$^{-1}$, FWHM(2D)$_{\text{1LG}}\sim$61cm$^{-1}$, FWHM(2D)$_{\text{2LG}}\sim$57cm$^{-1}$, FWHM(2D)$_{\text{3LG}}\sim$49cm$^{-1}$, FWHM(2D)$_{\text{4LG}}\sim$53cm$^{-1}$. Assuming 2686cm$^{-1}$ as unstrained Pos(2D)\cite{Mohiuddin2009} and a rate of change in Pos(2D) with strain $\delta$Pos(2D)/$\delta \epsilon\sim$-64cm$^{-1}$/\%\cite{Mohiuddin2009}, this would give$\sim$0.22\% uniaxial strain. This would also explain the rather large FWHM(2D)\cite{Mohiuddin2009,Yoon2011}. We assume the strain to be mostly uniaxial, because it is unlikely for it to be biaxial, i.e perfectly isotropic in all directions, unless this is induced on purpose such as e.g. in bubbles\cite{Zabel2012}. There is a small increase in I(D)/I(G) to$\sim$0.2 for 2-4LG on HDMs.
\subsection{Lubricant coating}
Commercial HDM with commercial COC and commercial PFPE lube are sourced from HGST\cite{Waltman2012}. We also coat all other samples with PFPE Zdol 4000 as follows. The lube is dissolved in Vertrel XF. The samples are then immersed into and withdrawn from the PFPE solution at$\sim$0.5mm/s. Before withdrawal the samples are held for 20s to allow the lubricant molecules to adhere. The PFPE-coated samples are then cured at 150$^{\circ}$C for 1.5h. This is done to improve the PFPE bonding to the surface\cite{Dwivedi2015}. The parameters are optimized to obtain a lubricant thickness$\sim$1.2$\pm$0.2nm\cite{Dwivedi2015,Dwivedi2016c,Waltman2012}, similar to commercial HDMs\cite{Waltman2012}. We also apply PFPE on BM (BML) to examine the tribology of lube containing bare HDM.
\subsection{Laser irradiation stability}
To test if SLG can withstand HAMR conditions we consider SLG transferred on L10 FePt-based HDM\cite{Richter2007,Plumer2001,Challener2009,Hussain2015} and on Si/SiO$_2$. Ref.\cite{Hussain2015} achieved T$_C$ at$\sim$1.3mW/$\mu$m$^2$ by optimizing aperture optics, Ref.\cite{Challener2009} suggested to use a laser power density$\sim$10$^7$W/cm$^2$ and$\sim$0.35$\times$10$^7$ W/cm$^2$ for FePt-based HDM. We perform Raman measurements at 785nm, the closest available to that used in HAMR ($\sim$830nm\cite{Challener2009,Hussain2015}), with a spot size$\sim$1.24$\mu$m, as determined by the razor blade technique\cite{razor-blade}. This involves using a razor blade between objective and sample such that it intersects the beam in a direction perpendicular to its propagation axis, and then moving it from fully covering the beam path until it does not, while measuring the peak intensity\cite{razor-blade}. If the profile of the beam is described by a Gaussian, the signal measured by the detector is represented by an integrated Gaussian\cite{razor-blade}, i.e. the derivative of the signal will give a Gaussian profile. We perform a line scan perpendicular to an Au contact, while measuring the Si peak intensity as function of scan position. We plot the Si peak intensity as a function of scan position. Taking the derivative of the intensity profile gives a Gaussian, and its 1/e$^2$ width corresponds to the laser spot diameter\cite{razor-blade}. We vary the power density from$\sim$1.3mW/$\mu$m$^2$ (0.013$\times$10$^7$W/cm$^2$) to$\sim$31.5mW/$\mu$m$^2$ (0.315$\times$10$^7$W/cm$^2$) in order to examine the laser irradiation-driven evolution of the SLG Raman spectrum under HAMR power densities.

We first consider Si/SiO$_2$/1LG. We record spectra at different power densities for$\sim$4mins to achieve a good signal to noise ratio. Fig.\ref{fig:fig2}c shows that 1LG on Si/SiO$_2$ has no D peak even for the highest power density. Pos(G) downshifts from$\sim$1588 to 1583cm$^{-1}$, indicating an increase in T$\sim$312.5K with respect to RT by taking $-0.016$cm$^{-1}$/K as shift of Pos(G) with T\cite{Calizo2007}. Fig.\ref{fig:fig2}d plots the data for 1LG on FePt HDM, with I(G) normalized to that at 1.3mW/$\mu$m$^2$. In this case, the FePt substrate is rotated at 4100rpm on a circular track with diameter$\sim$4mm to simulate rotating HDD conditions, with a 20mins acquisition time, much larger than the total laser irradiation time expected for a 5-years life of HAMR-based HDDs\cite{Ji2014}. No D peak is seen for all power densities, thus confirming the stability of 1LG.
\subsection{Friction and wear}
\begin{figure*}
\centerline{\includegraphics[width=170mm]{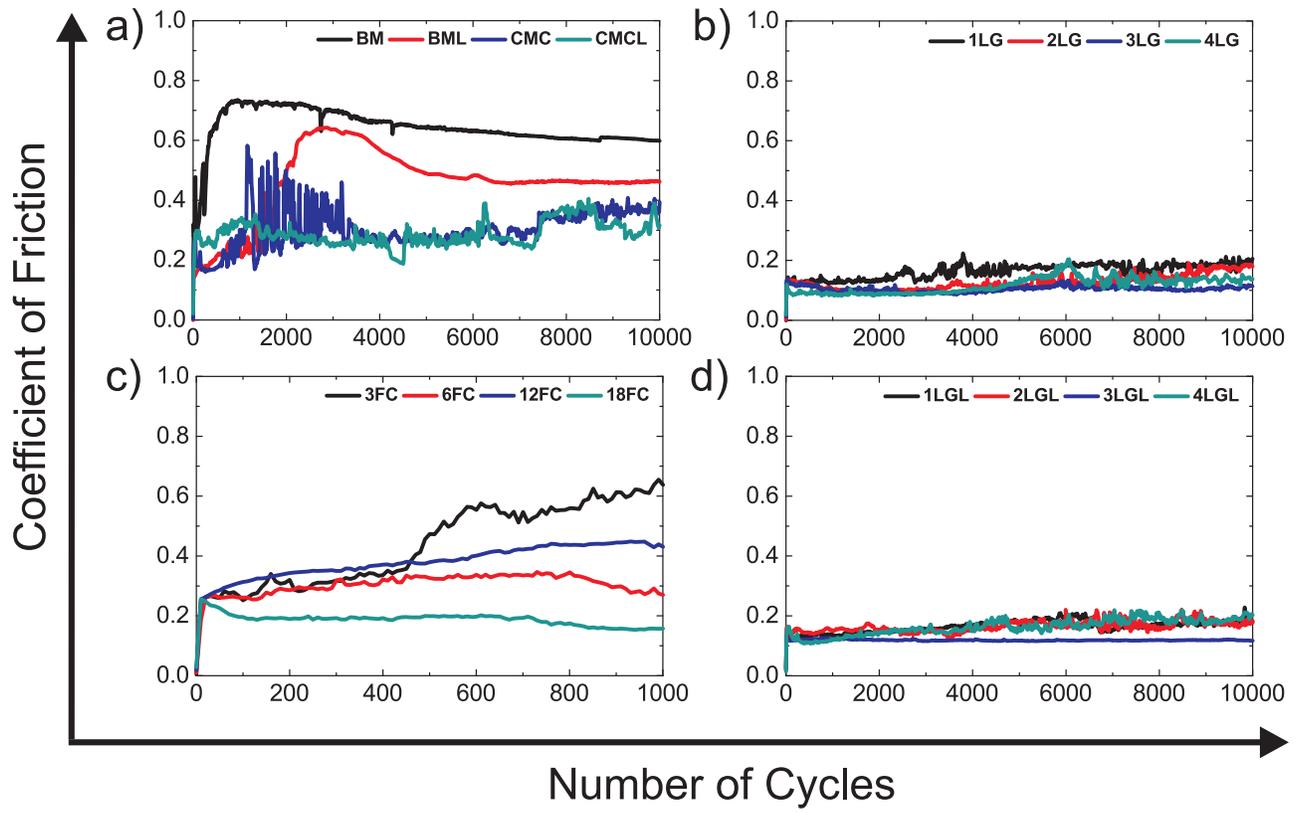}}
\caption{Representative friction curves for a) BM, commercial medium with commercial$\sim$2.7nm COC (CMC), commercial medium with commercial$\sim$2.7nm COC and$\sim$1.0nm PFPE lube (CMCL), BM coated with$\sim$1.2$\pm$0.2nm PFPE lube; b) HDM coated with 1-4LG; c) HDM with FCVA-ta-C of different thicknesses; d) HDM coated with 1-4LG and$\sim$1.2$\pm$0.2nm PFPE}
\label{fig:fig3}
\end{figure*}
\begin{figure*}
\centerline{\includegraphics[width=170mm]{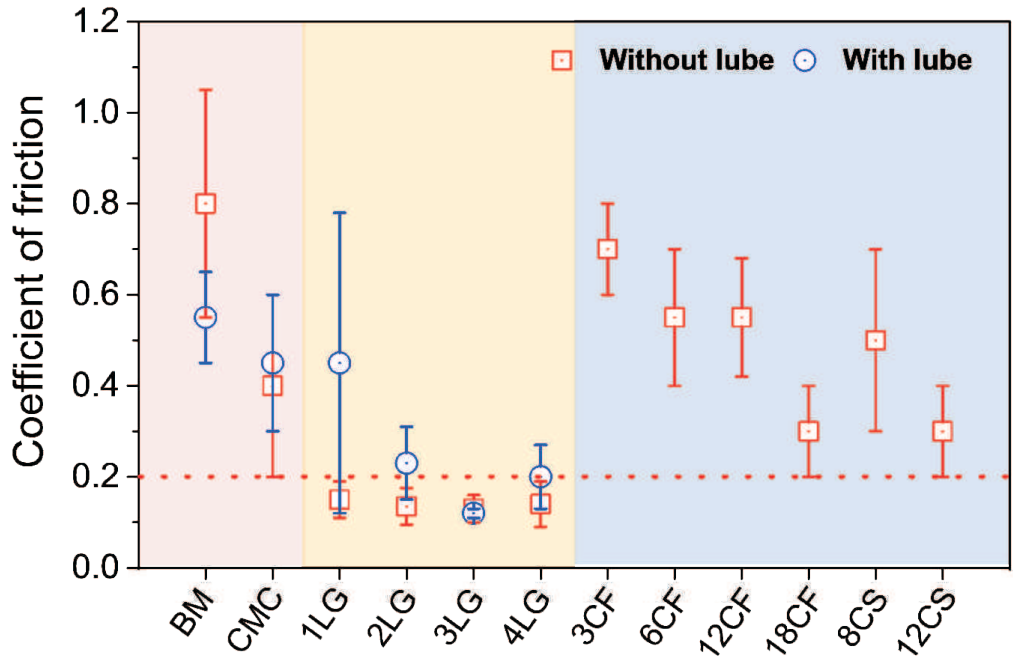}}
\caption{Average coefficient of friction with and without lubricant for 1-4LG and all COC reference samples}
\label{fig:fig3-2}
\end{figure*}
\begin{figure*}
\centerline{\includegraphics[width=170mm]{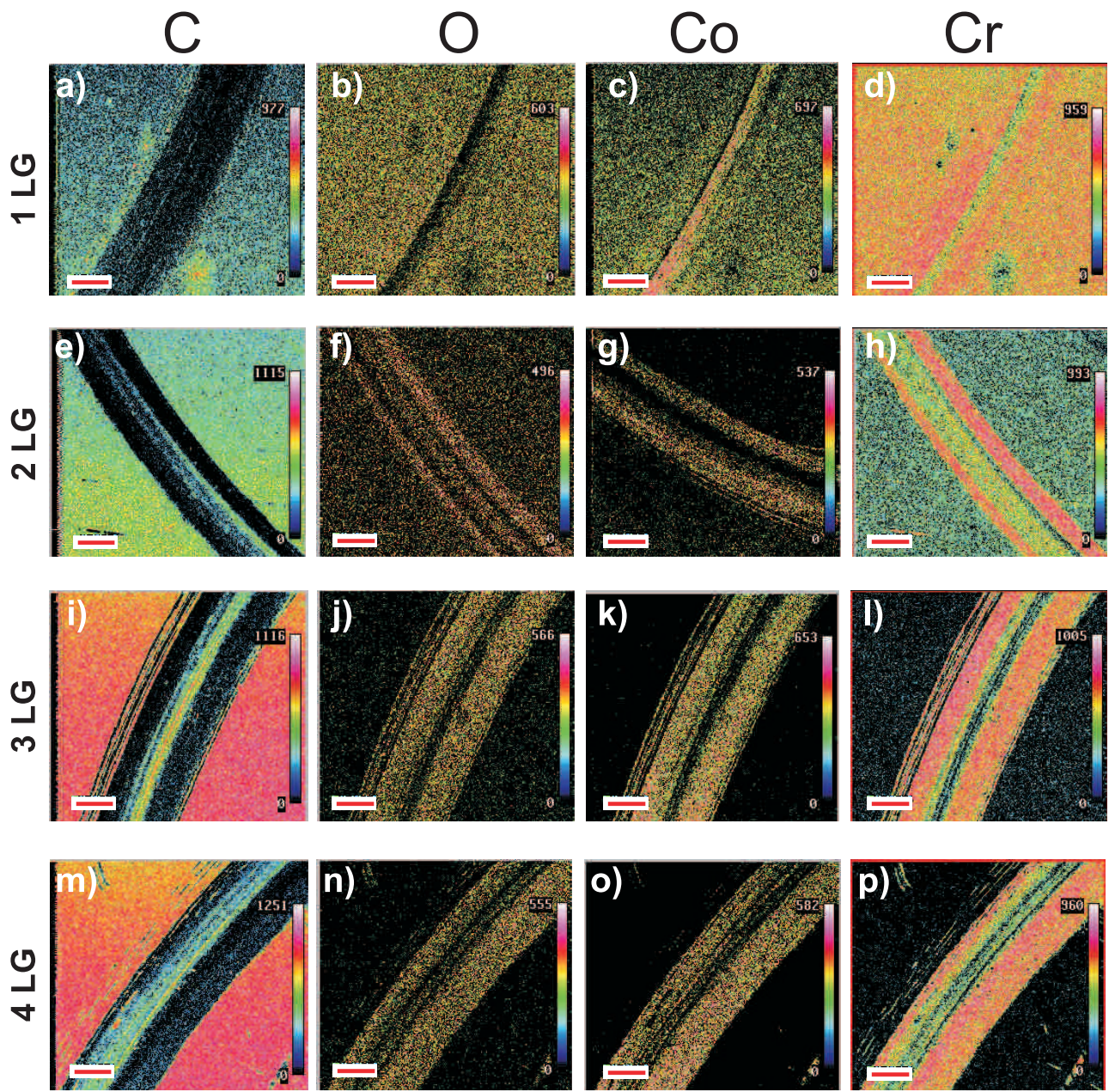}}
\caption{AES images of C, O, Co and Cr for a-d) 1LG, e-h) 2LG, i-l) 3LG and m-p) 4LG. Brighter colors indicate a higher intensity. Outside the wear track, the C signal (top layer) increases and Co and Cr signals (underlying substrate) decrease with increasing N due to limited AES sampling depth\cite{AES}}
\label{fig:fig4}
\end{figure*}
\begin{figure*}
\centerline{\includegraphics[width=180mm]{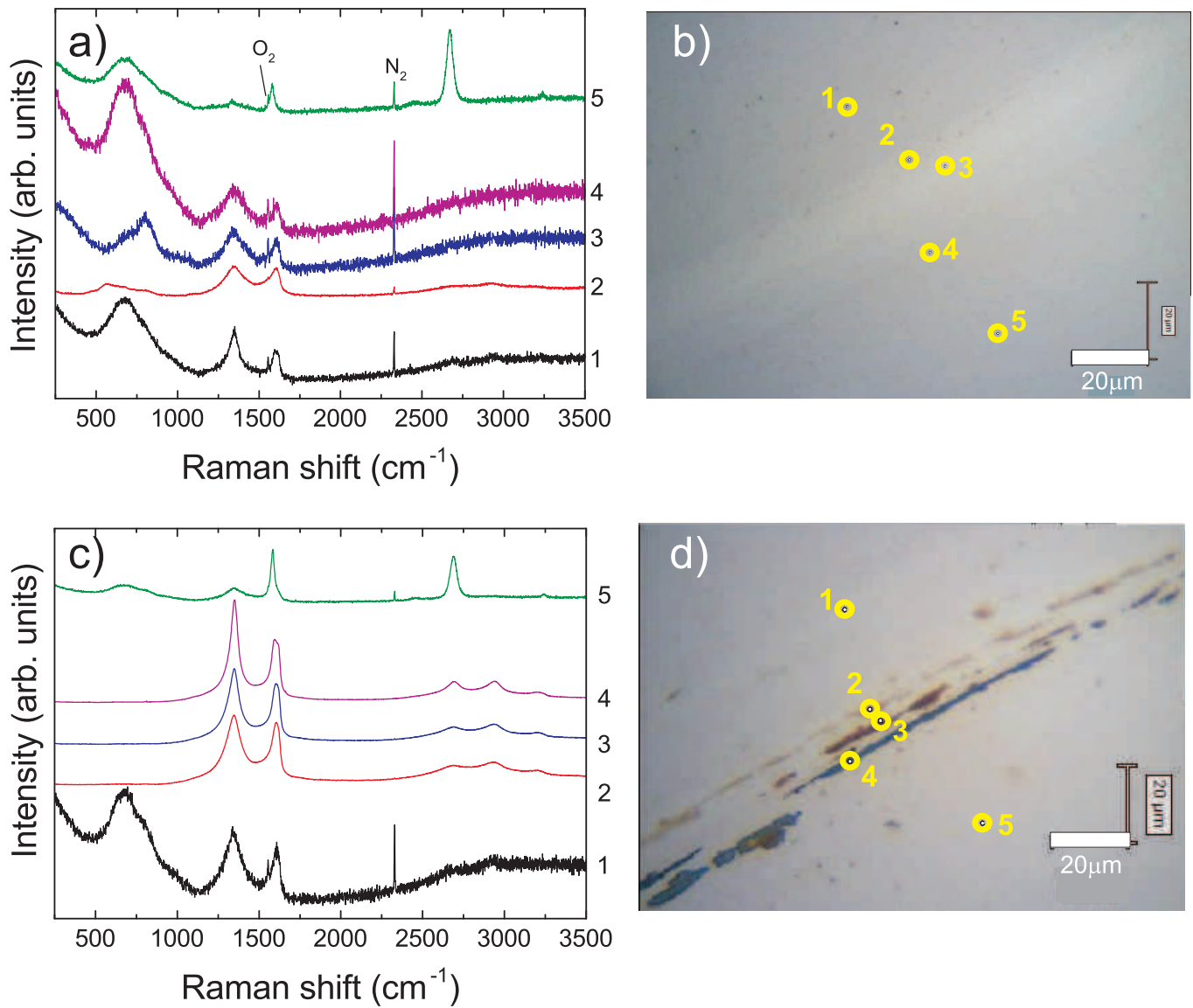}}
\caption{Raman spectra, normalized to I(G), across wear tracks of a) 1LG, c) 2LG and corresponding optical pictures in b),d)}
\label{fig:fig5-1}
\end{figure*}
\begin{figure*}
\centerline{\includegraphics[width=180mm]{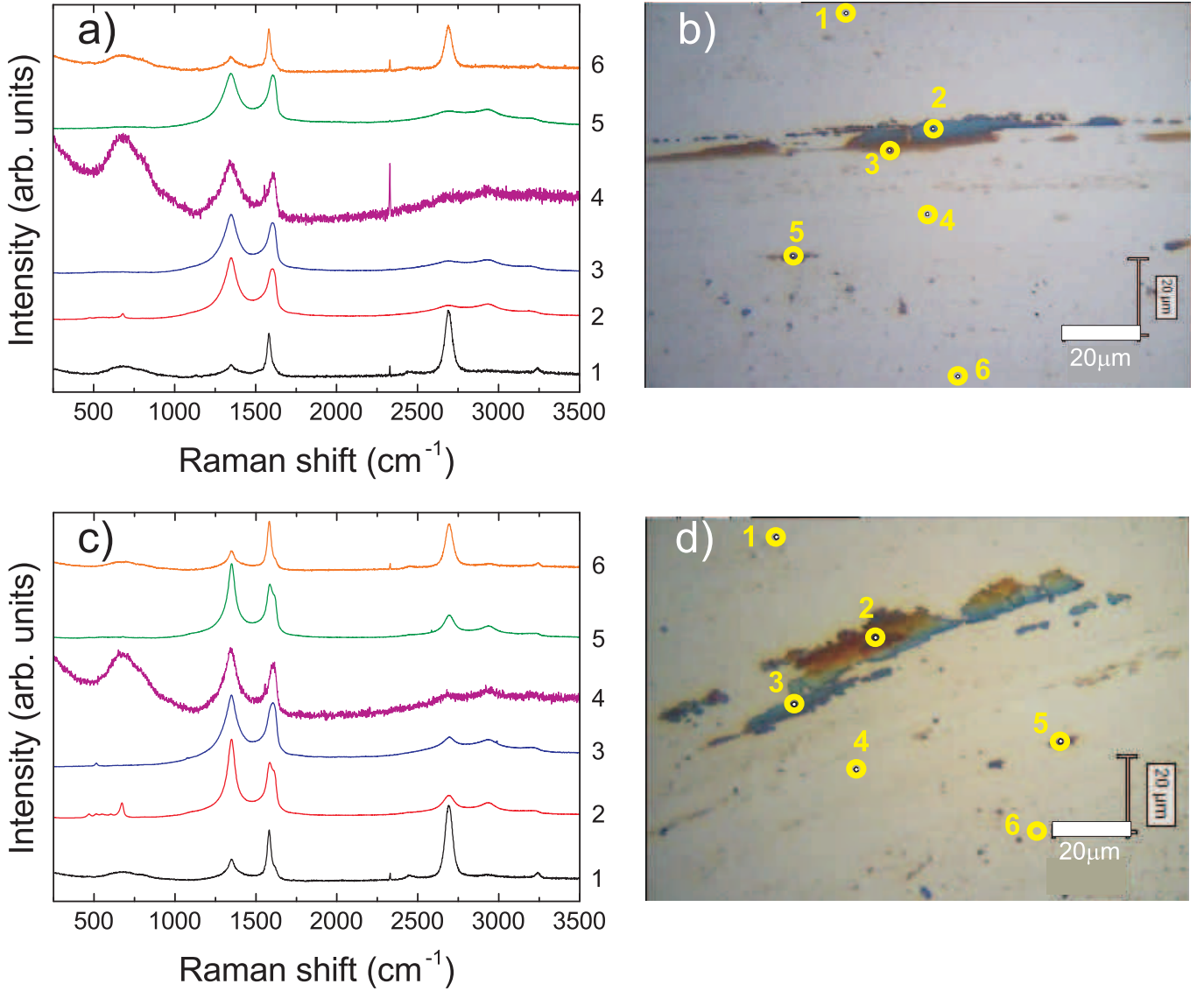}}
\caption{Raman spectra, normalized to I(G), across wear tracks of a) 3LG and c) 4LG and corresponding pictures in b,d)}
\label{fig:fig5-2}
\end{figure*}
\begin{figure}
\centerline{\includegraphics[width=85mm]{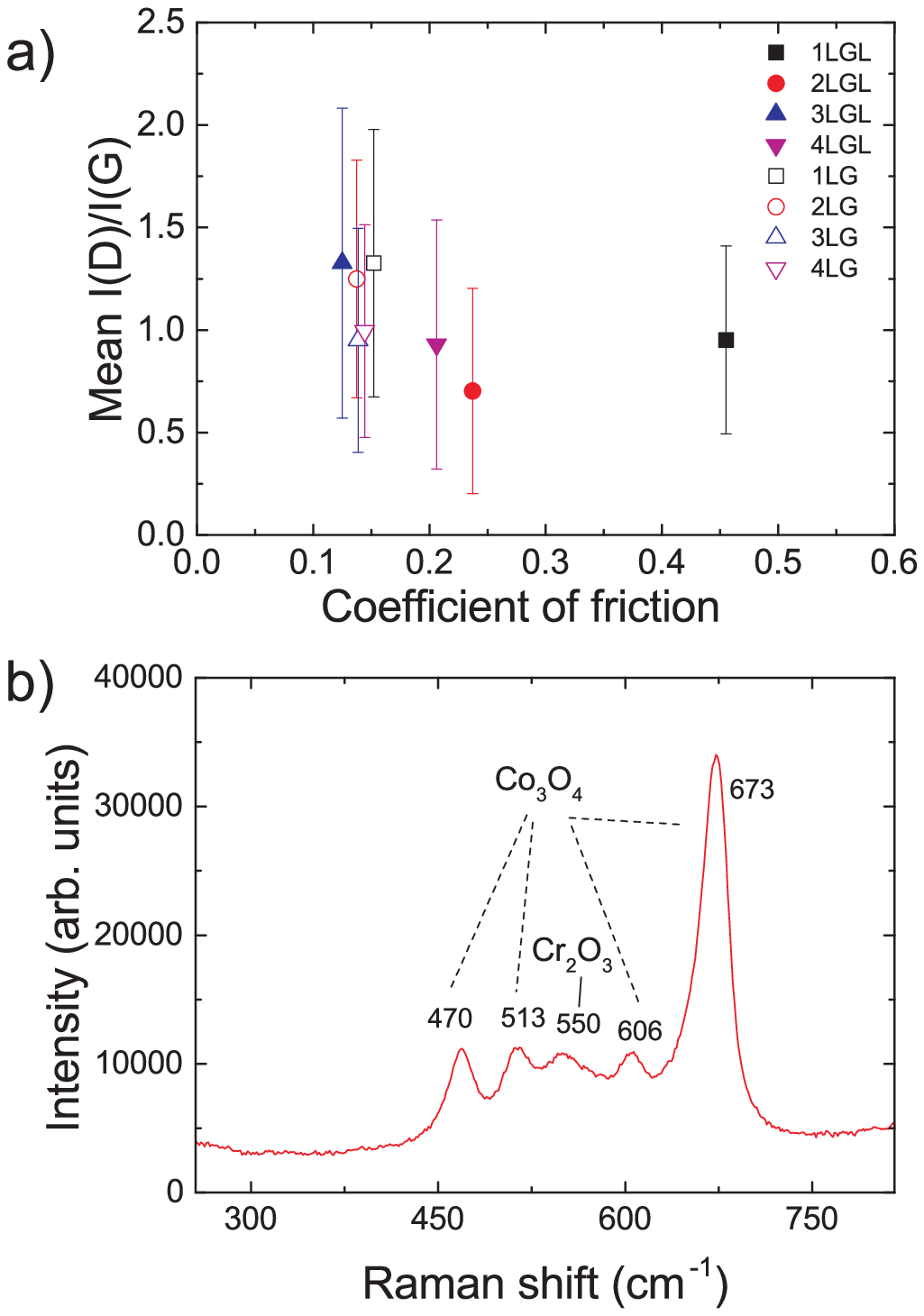}}
\caption{a) Mean I(D)/I(G) of 1-4LG coated samples with and without lube after friction tests as function of COF. b) Raman spectrum on a wear track. The region between$\sim$400-750cm$^{-1}$ shows peaks of the underlying substrate}
\label{fig:media-ID-IG}
\end{figure}

Since in HDDs the HDM spins during operation, friction and wear need to be examined in a setup mimicking the HDD operation. AFM and other tip-based tools are used to measure friction and wear\cite{Lee2010,Egbert2014}. In contrast to setups with rotating geometry, tip-based tools measure based on the movement of the tip in the lateral direction\cite{Lee2010,Egbert2014}. Ball-on-disk measurements (rotation based geometry) have a similar assembly as HDDs, with samples rotating while the counterface is in contact with the surface\cite{Dwivedi2015,Berman2015}, thus measuring friction. We perform ball-on-disk tests using a nano-tribometer (CSM Instruments) in a cleanroom to have controlled environment with T=23$\pm$1$^{\circ}$C and a relative humidity (RH)$\sim$55$\pm$5\%. A sapphire ball (Al$_2$O$_3$) of diameter$\sim$2.0$\pm$0.1mm and surface roughness$\sim$5.0$\pm$0.1nm is used as the counterface, because the HDH is made of an Al$_2$O$_3$-based composite\cite{10Marchon2013}. During the test, a normal load$\sim$20mN and a rotational speed$\sim$100rpm corresponding to a linear speed$\sim$2.1cm/s for 10000 cycles are used. Since in HDDs the contact occurs occasionally\cite{10Marchon2013}, 10000 cycles in our setup are much higher than the operational lifetime of HDDs. After each test, the wear track and ball images are captured using an optical microscope. To check repeatability, tests are performed 2-7 times.

Figs.\ref{fig:fig3}a-c plot representative friction curves for BM, CMC, CMCL and BML (Fig.\ref{fig:fig3}a) coated with 1-4LG (Fig.\ref{fig:fig3}b), and media coated by FCVA (Fig.\ref{fig:fig3}c). The average COFs are in Fig.\ref{fig:fig3-2}, including media coated by DC-sputtering. BM has the highest COF$\sim$0.8, with substantial wear as confirmed by optical images of balls and wear tracks in Figs.\ref{fig:fig5-1},\ref{fig:fig5-2}. The COF of$\sim$2.7nm CMC reaches$\sim$0.4 at 10000 cycles, but with strong fluctuations between$\sim$0.2 and 0.6 at 1500 up to $\sim$3500 cycles, Fig.\ref{fig:fig3}a, and negligibly improves wear with respect to BM, Fig.\ref{fig:fig3}a. The transfer of 1-4LG reduces COF below 0.2 for all samples, and gives non-fluctuating, smooth friction curves. Furthermore, the COF for 1-4LG-coated samples (without lube) is$\sim$4 times lower than BM and$\sim$2 times lower than CMC, despite a reduction of thicknesses$\sim$7 times (for 1LG) to 2 times (for 4LG) with respect to CMC. Figs.\ref{fig:fig5-1}b,d,\ref{fig:fig5-2}b,d reveal that the wear track width of 1-4LG-coated samples is$\sim$2-4 times lower and debris transferred to the ball are smaller than BM and CMC, indicating higher wear resistance. All FCVA COCs with thicknesses from$\sim$0.3 to$\sim$1.8nm show$\sim$2 to 5 times higher COF than 1-4LG-coated samples with and without lube, apart from 1LG without lube. Pulsed DC sputtered COCs have$\sim$2-3 times higher COF than 1-4LG-coated samples.

To further analyze the wear tracks, Auger Electron Spectroscopy (AES) imaging is performed using a JEOL JAMP Auger Microprobe, Fig.\ref{fig:fig4}. Before recording AES images, scanning electron microscope (SEM) images are taken to select the AES locations. The AES images inside and outside the wear tracks show the carbon-containing sites and that the amount of carbon on the wear track increases with increasing number of graphene layers N. The Co and Cr intensities inside the wear tracks are higher for 1LG, and decrease with increasing N, due to the increase in C and$<$1-3nm sampling depth of AES\cite{AES, AES1,AES2}. The O signal in the wear track appears due to ambient oxygen as the samples are exposed to air before AES, with some contribution from the media oxide.

After ball-on-disk tests, the wear tracks are analyzed by Raman spectroscopy. Ref.\cite{Ferrari2000} introduced a three stage model of amorphization. Stage 1: graphene to nanocrystalline graphene. Stage 2: nanocrystalline graphene to low sp$^3$ amorphous carbon. Stage 3: low sp$^3$ amorphous carbon to high sp$^3$ amorphous carbon. For all spectra in Figs.\ref{fig:fig5-1},\ref{fig:fig5-2}, as the D peak increases, D' and D+D' appear, whereas the 2D peak intensity weakens when approaching the wear track, indicating an increase in disorder according to stage 1. The broad peak between 500 and 1000cm$^{-1}$ is due to the glass substrate\cite{White1984,Affatigato}. At the centre of the wear track, all second order Raman features (i.e. 2D, 2D' and D+D' peaks) merge, while I(D) decreases and D,G become broader, indicating an increase of disorder\cite{Ferrari2000}. 2-4LG are less damaged than SLG, as for the spectra in Figs.\ref{fig:fig5-1}a,c and I(D)/I(G) in Fig.\ref{fig:media-ID-IG}a, from at least 5 positions as a function of COF without and with lubricant (L). Overall, there seems to be no significant trend of I(D)/I(G) with COF. However, a difference is seen without lubricant: I(D)/I(G) is 0.95 and 1.0 for 3-4LG, while for 1-2LG it is 1.33 and 1.25. Thus, there are less defects in 3 and 4LG compared to 1 and 2LG. The peaks below$\sim$1000cm$^{-1}$ in Fig.\ref{fig:media-ID-IG}a are due to the underlying HDM: Those$\sim$470cm$^{-1}$ (E$_g$), 513cm$^{-1}$(F$_{2g}$), 606cm$^{-1}$(F$_{2g}$), 673cm$^{-1}$ (A$_{1g}$) are from Co$_3$O$_4$\cite{Gallant2006,Hadjiev1988}, while the A$_{g}$ mode$\sim$550cm$^{-1}$ is from Cr$_2$O$_3$\cite{Banerjee2014}.
\subsection{Effect of lubricant}
Friction measurements on lube-coated samples (Figs.\ref{fig:fig3}d,e) show that the COF of 2-4LG is similar to their non-lubricated counterparts, suggesting that 2-4LG without lube are lubricious enough and that the lube does not improve lubricity. Lube on 1LG results in higher and inconsistent friction and wear, as some measurements show lower COF$\sim$0.15, some higher$\sim$0.78, with friction increasing after few thousands cycles. The COF of 1LG, on an average, is$\sim$0.45, i.e.$\sim$3 times its non-lubricated counterpart. The D and G peaks in the Raman measurements in Figs.\ref{fig:fig5-1},\ref{fig:fig5-2} on 1-4LG confirm the presence of carbon on the wear tracks and transfer of debris to the balls, similar to the non-lubricated case.
\subsection{Corrosion}
Co-alloys have great propensity to corrode, mainly due to Co oxidation\cite{Tomcik2000}. This results in the loss of magnetic properties\cite{Tomcik2000}, giving one of the major concerns for long-term functionality and durability of HDDs\cite{Casiraghi2007,Dwivedi2015c,Tomcik2000}. To examine the corrosion protection efficiency of 1-4LG and compare their performance with state-of-the-art COCs, the corrosion of different uncoated- and coated-HDM, exposed to an electrolyte solution$\sim$0.1M NaCl similar to that used in Refs.\cite{Tomcik2000} on a$\sim$0.24cm$^2$ area, is investigated using an electrochemical corrosion method\cite{Dwivedi2016c,Tomcik2000}. The measurements are performed with a 3-electrode setup with a Pt wire as counter electrode, Ag/AgCl as reference electrode and HDM as working electrode, to which the potential is applied\cite{Dwivedi2015c}. Each test consists of anodic and cathodic sweeps, where the potential is varied and the corresponding current measured\cite{Dwivedi2015c}. Every sweep is conducted at different locations, with at least 3 sets of 6 sweeps on each sample. The so-called Tafel's analysis\cite{Tomcik2000, Stansbury} is done by plotting anodic and cathodic curves on a semi-logarithmic scale of potential versus log current. The linear part of the logarithmic anodic and cathodic currents are extrapolated\cite{Tomcik2000,Stansbury}, and the intercept of these lines gives the corrosion current\cite{Tomcik2000,Stansbury}, and the corrosion current density J$_{\text{corr}}$ when divided by the contact area. The corrosion protection efficiency (CPE), a parameter which defines how efficient the overcoat is in protecting against HDM corrosion, is estimated as\cite{Dwivedi2016c}:
\begin{equation}
CPE(\%)=\frac{J^0_{\text{corr}}-J_{\text{corr}}}{J^0_{\text{corr}}}\times 100,
\end{equation}
where $J^0_{\text{corr}}$ is the corrosion current density of BM and J$_{\text{corr}}$ that of coated HDM.

\begin{figure}
\centerline{\includegraphics[width=90mm]{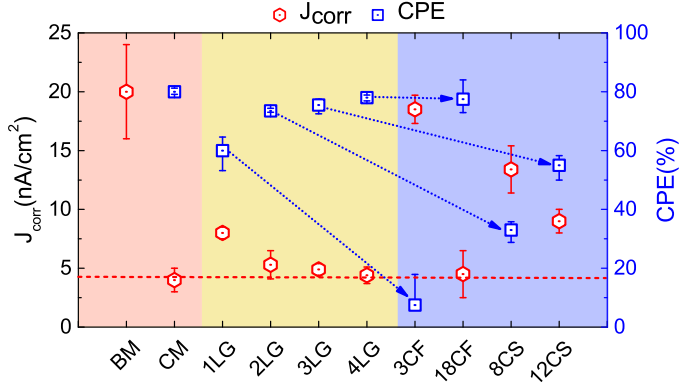}}
\caption{$J_{\text{corr}}$ and CPE for different samples. The corrosion resistance varies inversely with J$_{\text{corr}}$\cite{Dwivedi2015c}. 2-4LG have similar CPE to$\sim$2.7nm CMC and$\sim$1.8nm amorphous carbon in sample 18CF. The dotted line is a guide to the eye. The arrow compares various 1-4LG-COCs pairs, such as 1LG$\rightarrow$3CF, 2LG$\rightarrow$8CS, 3LG$\rightarrow$12CS and 4LG$\rightarrow$18CF, as the 1-4LG sample in each pair is always slightly thinner than the COC.}
\label{fig:fig7}
\end{figure}

When a metal is exposed to a corrosive solution, it releases ions that leave behind electrons, which can be observed in an anodic reaction\cite{Tomcik2000} as:
\begin{equation}
M\rightarrow M^{n+}+ne^{-},
\end{equation}
where M is the metal and n the number of electrons released. For Co-alloy-based systems, metal dissolution, decreasing anode conductivity\cite{Tomcik2000}, can take place\cite{Tomcik2000}:
\begin{equation}
Co\rightarrow Co^{2+}+2e^{-},
\end{equation}
As the corrosion reaction involves the transfer of electrons and ions between metal and solution\cite{Tomcik2000}, the corrosion rate varies with corrosion current\cite{Tomcik2000}, hence J$_{\text{corr}}$ varies inversely with corrosion resistance, i.e. the COC ability to reduce the HDM corrosion\cite{Dwivedi2016c,Dwivedi2015c,Tomcik2000}. The BM shows the highest J$_{\text{corr}}$, indicating the greatest propensity to corrode, Fig.\ref{fig:fig7}. J$_{\text{corr}}$ decreases and CPE increases with increasing N. 1LG reduces J$_{\text{corr}}\sim$2.5 times with respect to BM, but it is twice that of HDM with $\sim$2.7nm commercial COC. For 2LG, J$_{\text{corr}}$ is$\sim$3.8 times smaller than BM, and comparable to CMC. This is remarkable as the 2LG thickness$\sim$0.7nm, is$\sim$4 times lower than CMC with$\sim$2.7nm COC. J$_{\text{corr}}$ then marginally reduces for increasing N. Fig.\ref{fig:fig7} shows that, for similar thickness, graphene-based overcoats have lower J$_{\text{corr}}$ than amorphous COCs, indicating greater protection. The CPE of 2-4LG is similar to thicker commercial CMC and higher than amorphous COCs of similar thickness.

The COC defects and pinholes are the active corrosion sites\cite{Tomcik2000}, and their conductivity could further add galvanic corrosion\cite{Tomcik2000}. The COC should be defect and pinhole-free, smooth and with excellent barrier properties to minimize HDM oxidation and corrosion. Ref.\cite{Prasai2012} showed that 1LG can be used as corrosion barrier for metallic surfaces provided it is uniform and defect-free. In our case 1LG reduces corrosion$\sim$2.5 times with respect to BM. 2LG results in$\sim$3.8-fold reduction in J$_{\text{corr}}$ with respect to BM, and its J$_{\text{corr}}$ approaches that of$\sim$4times thicker commercial COC. At and beyond 2LG, J$_{\text{corr}}$ and CPE remain similar, adding marginal anti-corrosion improvement, Fig.\ref{fig:fig7}. The higher corrosion protection in 2-4LG with respect to 1LG is mainly due to the fact that the increase in N improves the coverage of HDM, leading to reduction of active corrosion sites.
\section{Discussion}
Friction is defined as resistance to sliding\cite{Berman2014,Bowden1942,Postnikov1964}. At the macroscopic level, Amonton's law\cite{Bowden1942,Postnikov1964} states that the frictional force between two bodies varies proportionally to the normal force. Hence:
\begin{equation}
COF=\frac{F}{W},
\label{eq:mu}
\end{equation}
where F[mN] is the frictional force and W[mN] is the normal force.

Amonton's law does not take into account the area of contact at the microscopic level\cite{Bowden1942}. Ref.\cite{Bowden1942} suggested that the contact between two bodies contains several smaller contacts, called asperities, with the sum of the area of these asperities being lower than the apparent macroscopic area\cite{Bowden1942}. Thus, from Ref.\cite{Bowden1942}, F in micro-tribology can be expressed as:
\begin{equation}
F=\tau \sum A_{asp},
\label{eq:F}
\end{equation}
where $\tau$ is the shear strength, i.e. the stress required to shear the contacting interfaces and enable sliding\cite{friction-book2}, expressed as shear force/area and $\sum A_{asp}$ is the sum of the asperities areas, also called real contact area\cite{Bowden1942}. Hence, friction also depends on $\sum A_{asp}$. Materials with lower strength than the contacting bodies undergo plastic deformation\cite{Bowden1942}, which depends on W\cite{Bowden1942}:
\begin{equation}
W=p \sum A_{asp},
\label{eq:N}
\end{equation}
where $p$ [N/m$^2$] is the flow pressure, i.e. the ratio of force required to displace a material, e.g. a metal, from the slides in a friction measurement setup divided by the cross-section of the torn track\cite{Bowden1942}. Thus, Eqs.\ref{eq:mu},\ref{eq:F},\ref{eq:N}, give:
\begin{equation}
COF=\frac{F}{W}=\frac{\tau}{p}
\label{eq:mu-final}
\end{equation}
A Co-alloy is mechanically softer\cite{Liu2002} than Al$_2$O$_3$\cite{Reidl2012}. When a metallic Co-alloy slides against Al$_2$O$_3$, it causes the HDM surface to undergo plastic deformation\cite{Bowden1942,Postnikov1964}, generating wear and high COF$\sim$0.8, Figs.\ref{fig:fig3},\ref{fig:fig3-2}. This is consistent with the friction results for metallic surfaces in Refs.\cite{Bowden1942,Postnikov1964}. This could be due to the formation of an adhesive contact at the HDM-ball interface enabled by the high surface energy of Co-alloy-based HDM$\sim$42.8mN/m\cite{Rana2016} and the presence of contaminants at the interface that may enhance the interaction between the two bodies\cite{Li2016}. When 1LG is placed on HDM, the COF decreases$\sim$4 times and negligibly changes with N.
\begin{figure*}
\centerline{\includegraphics[width=170mm]{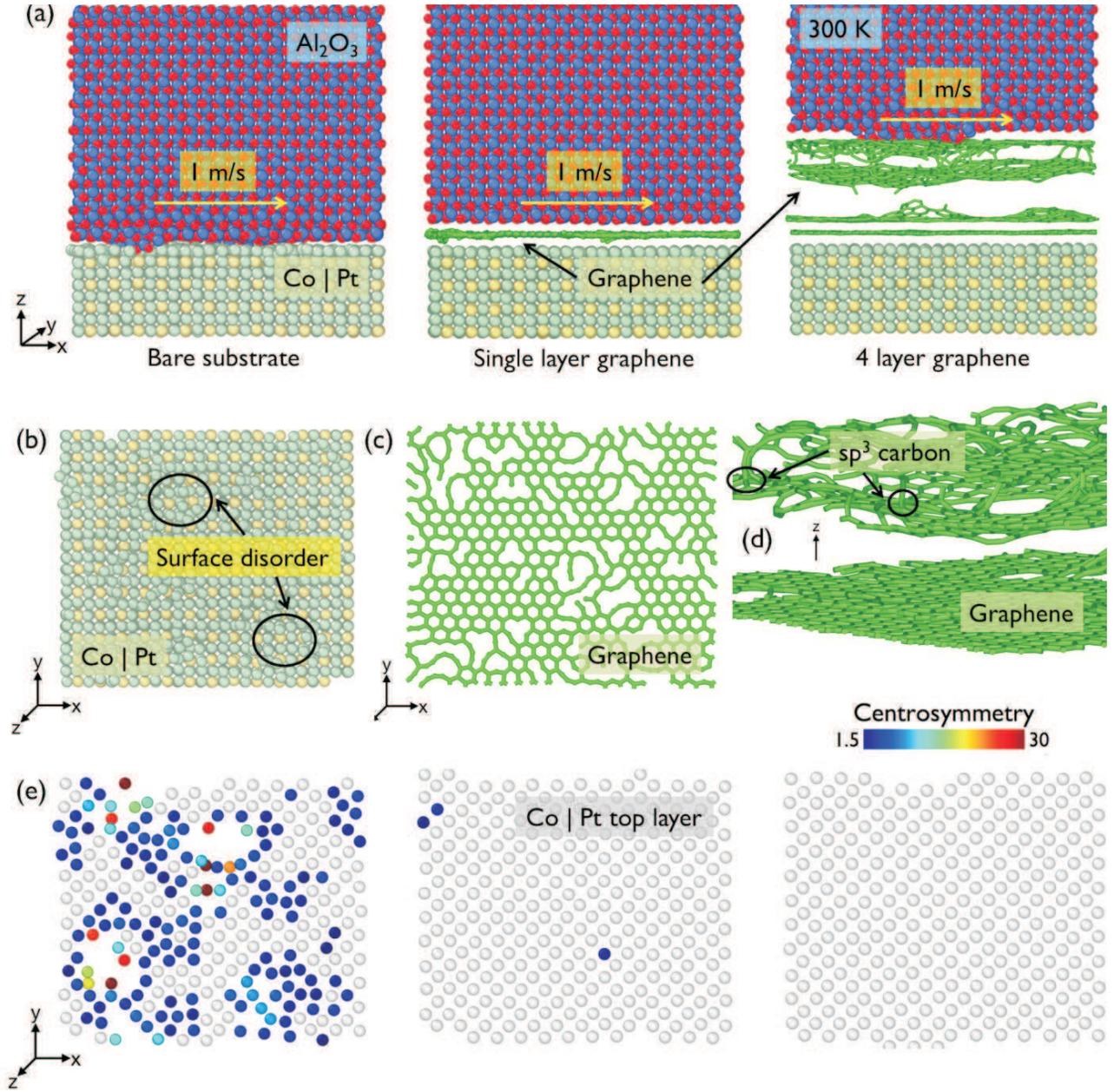}}
\caption{a) Simulation setup for evaluating wear between Al$_2$O$_3$, Co|Pt,  1 and 4LG. T=300K and the materials slide against each other at 1m/s. b) Interfacial surface atoms of bare Co|Pt showing disorder. Two representative disorder regions are indicated by circles. c) Structure of defected 1LG. d) Structure of 4LG. The top two layers are attracted towards the Al$_2$O$_3$ block with the formation of sp$^3$ crosslinks. e) Disorder of interfacial Co|Pt surface using the centrosymmetry parameter.}
\label{fig:fig8}
\end{figure*}
\begin{figure*}
\centerline{\includegraphics[width=170mm]{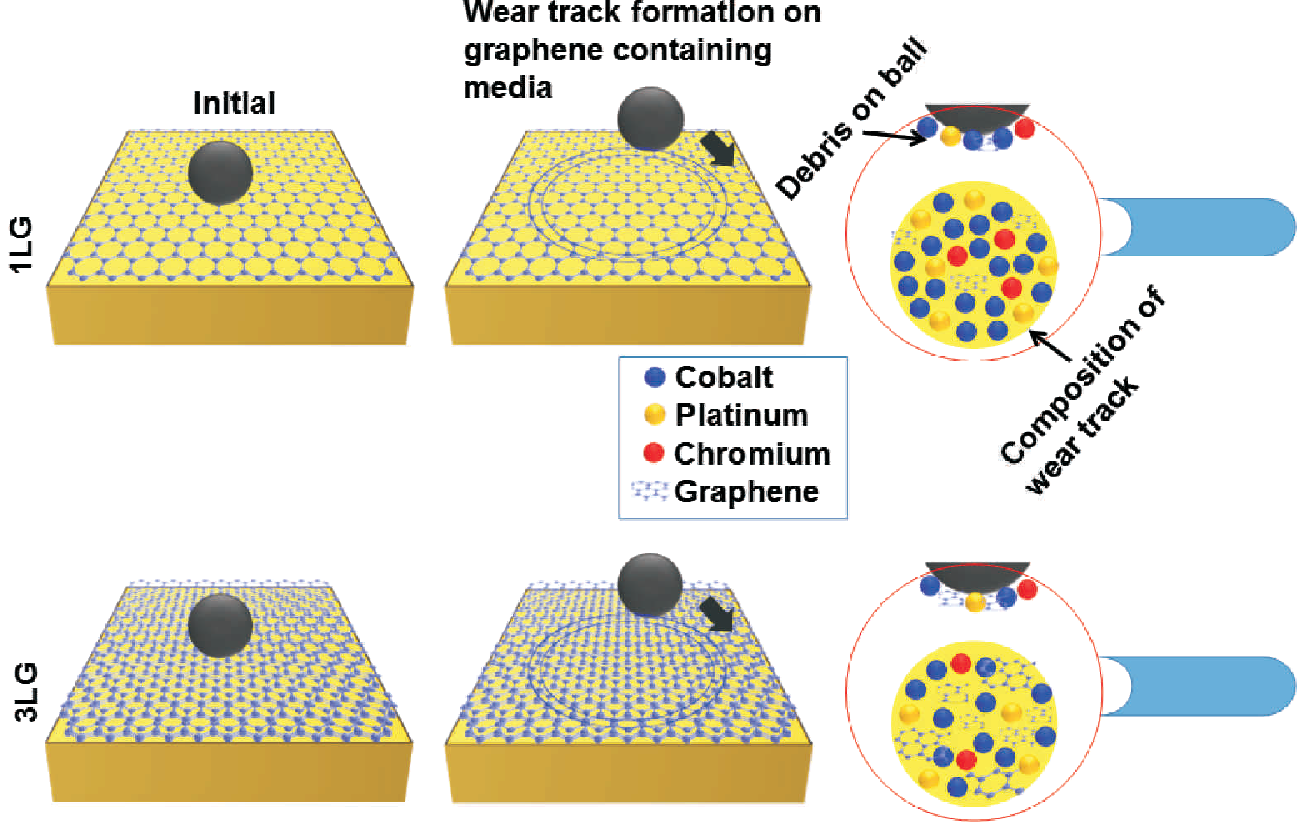}}
\caption{Interaction of counterface ball (sapphire) with 1 and 3LG. The zoomed-in images represent 1 and 3LG, and underlying substrate atoms (Co, Pt, Cr, etc.) on the counterface ball and wear track where the amount of carbon debris increases with N. The debris generated on both interacting bodies maintain the lower COF of 1-4LG with respect to BM.}
\label{fig:fig6}
\end{figure*}

At the nanoscale, the friction of 1LG on Si/SiO$_2$ against Si tips, measured at a low load$\sim$1nN and scan speed$\sim$1-10$\mu$m/s, was explained based on out-of-plane deformation in front of a scanning probe tip, the so-called puckering effect\cite{Lee2010,Li2010}. This enhances the contact area, hence friction. The out-of-plane deformation is suppressed with increasing N\cite{Li2010}, resulting in reduced COF. This theory is not applicable to 1-4LG, due to the significantly larger dimension (ball diameter$\sim$2.0$\pm$0.1mm) of the counterface, much higher load$\sim$20mN and higher speed$\sim$100rpm or 2.1cm/s, which do not differentiate the frictional characteristics of 1-4LG. Thus, all coated media show similar friction, Fig.\ref{fig:fig3}b.

1LG is mechanically strong\cite{Lee2008,Lee2013}. It reduces the surface energy of various surfaces\cite{Wang2009,Shin2010,Rafiee2012} and can lead to incommensurate tribological contacts\cite{Berman2015} between magnetic media and counter-face. Thus, compared to BM, the COF reduction of 1-4LG-coated media could be attributed to 1LG's mechanical properties: breaking strength$\sim$42 Nm$^{-1}$\cite{Lee2008}, Young's Modulus$\sim$1TPa\cite{Lee2008}, flexibility (1LG can be stretched up to$\sim$20\% without breaking\cite{Lee2008,Bunch2007}), reduced adhesive interaction (lower surface energy)\cite{Lee2013,Wang2009,Shin2010,Rafiee2012}, as well as incommensurability of the lattice planes sliding against each other at the tribological interface\cite{Berman2015}, i.e. the hills of one surface with lattice spacing a do not match the valleys of the other surface of lattice spacing b, such that the ratio b/a is an irrational number\cite{incommensurate}. This is consistent with what suggested in Ref.\cite{Berman2014b} for 3-4LG on Si/SiO$_2$ sliding against an a-C:H-coated steel ball, and for 1, 3-4LG on a steel substrate sliding against a steel ball\cite{Berman2014b} in macroscale ball-on-disk conditions, with a normal force in z direction N$_z\sim$0.5$-$3N and speeds$\sim$0.6$-$25cm/s\cite{Berman2015, Berman2014b}. Since graphene layers are coupled by van-der-Waals forces, they shear easily\cite{Daly2016,Berman2013d, Wang2017,Ruiza2015}, as their interfacial shear strength is low, with a shear force per unit area$\alpha$=12.8$\cdot$10$^{18}$Nm$^{-3}$\cite{Tan2012} and a resulting shear modulus of C$_{44}$=4.3GPa\cite{Tan2012}. The ease of shearing further aids the reduction of friction in 2-4LG.

The AES on the wear tracks and Raman measurements across the wear tracks, after ball-on-disk tribological tests, in Figs.\ref{fig:fig4},\ref{fig:fig5-1},\ref{fig:fig5-2} show that a carbon signal is still present in all 1-4LG samples, even though disorder-induced peaks appear. Refs.\cite{Marchetto2012,Marchetto2015} also found carbon on wear tracks after micro-scale friction and wear tests of graphene on SiC. The counterface also reveals transferred debris, consisting of disordered carbon and underlying substrate atoms. This implies that when 20mN is applied and the sample rotates at$\sim$2.1cm/s, 1LG could turned into patches, due to the large,$\sim$20mN, contact load. Refs.\cite{Marchetto2012,Marchetto2015} reported transformation of continuous layers of 1-2LG into 1 and 2LG patches during microscale tribology, despite using a lower load$\sim$0.1$-$1mN and lower speeds$\sim$30-50$\mu$m/s than us. This leads to a distribution of 1LG patches along the wear track and transfer of 1LG-containing debris on the counterface. This also happens for 2-4LG coated media. Raman measurements in Figs.\ref{fig:fig5-1},\ref{fig:fig5-2} show disordered carbon on the wear tracks, with disorder lower than 1LG, as revealed by I(D)/I(G), Fig.\ref{fig:media-ID-IG}. This implies that 1LG acts as lubricant when in contact with another 1LG and is not completely removed during tribological tests. AES also reveals a progressive increase in the amount of carbon and carbon-containing sites on the wear tracks with increasing N. The debris transferred to the ball also contain carbon. Therefore, the formation of disordered carbon debris on both surfaces facilitates smooth sliding, and contributes to maintaining the low COF. The marginally lower friction in 2-4LG as compared to 1LG can be linked to I(D)/I(G), Fig.\ref{fig:media-ID-IG}a, with lower disorder corresponding to lower friction, although this does not apply for lube-containing all-1-4LGL samples. Fig.\ref{fig:fig6} schematically represents the proposed mechanism for friction reduction for 1 and 3LG-coated HDM, where carbon debris are on both ball and wear track.
\begin{figure}
\centerline{\includegraphics[width=90mm]{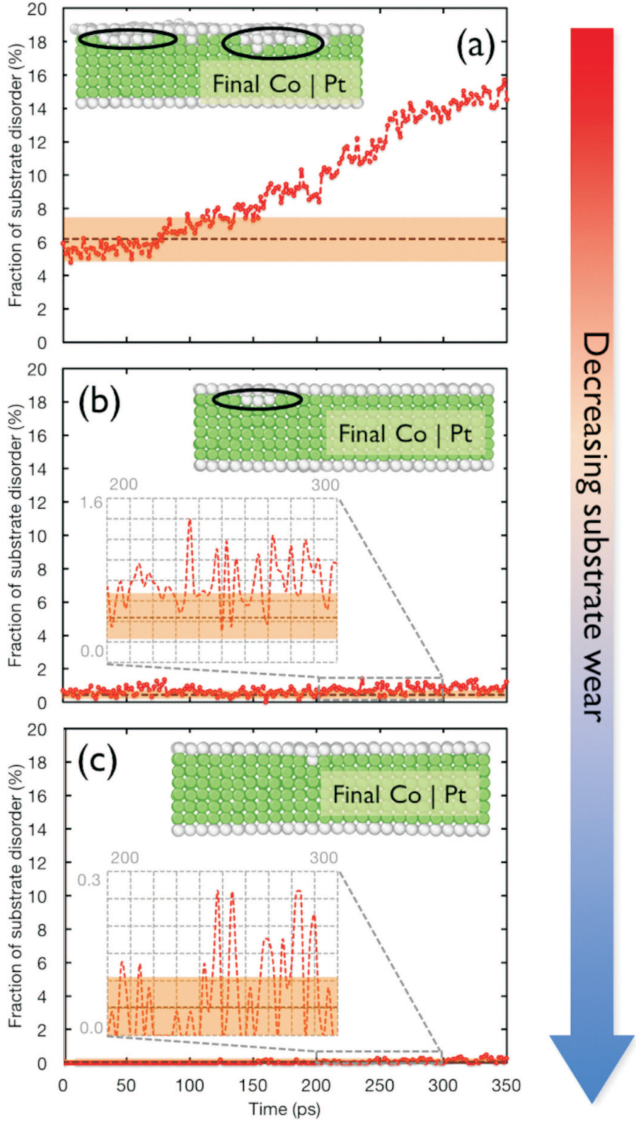}}
\caption{Fraction of Co|Pt disorder for a) HDM, b) HDM+1LG, c) HDM+4LG. In (a-c) the horizontal dotted line represents the HDM disorder before friction at 300K by virtue of cross interactions between alumina, 1/4LG and substrate. The transparent orange box indicates the standard error in the fractional substrate disorder at 300K. The insets shows atomistic snapshots of Co|Pt at 350ps. The green atoms have a local FCC environment, while the white are surface and/or defects. The secondary inset in panels (b-c) report the variation of fractional disorder from 200-300ps.}
\label{fig:fig9}
\end{figure}

To validate these assumptions, we perform molecular dynamics (MD) simulations of a Co|Pt$-$sapphire system with and without 1 and 4LG. Since Co-alloy-based HDM contain$\sim$70-75\% Co and$\sim$15-20\% Pt (total Co|Pt$\sim$90-95\%)\cite{Piramanayagam,Dwivedi2016c,Tomcik2000}, we use Co|Pt to mimic our Co-alloy-based HDM. The simulations are performed using the reactive force field (ReaxFF)\cite{Duin2001,Zhang2004} interatomic potential in order to describe the interactions between C in 1 and 4LG, Al and O in sapphire. The potential parameters for C/Al/O interactions are obtained from Refs.\citenum{Sen2013,Sen2014}. The Co|Pt substrate is assumed to be a fixed wall. This is reasonable, as the goal is to demonstrate that the honeycomb structure of 1 and 4LG, as interfacial layer between AlO$_x$ and Co|Pt, does not undergo significant changes, i.e. structural disorder, due to friction at 300K, even with a sliding velocity$\sim$1m/s. In addition, no chemical reactions occur between 1 or 4LG and AlO$_x$. This indicates that it is sufficient to use van-der-Waals interactions to describe the cross-interactions between different elements. For more efficient simulations, the cross-interaction across the Al$_2$O$_3$|1LG|Co$-$Pt interface is described by the 12$-$6 Lennard-Jones (LJ) potential\cite{Agrawal2002}. The interactions in 1 and 4LG are described by the bond-order Tersoff interatomic potential\cite{Tersoff1988}. The final MD simulations are performed in a large-scale atomic/molecular massively parallel simulator (LAMMPS)\cite{Plimpton1995} with hybrid/overlay pairs. AlO$_x$ is described by the embedded-atom method (EAM)\cite{Zhou2004} parameters developed as part of the charge-transfer ionic potential (CTIP) formalism\cite{Zhou2004,Streitz1994}. The interactions in 1 and 4LG are described by the Tersoff potential of Ref.\citenum{Lindsay2010}, while the Co|Pt substrate is described by the EAM potential of Ref.\citenum{Zhou2004}. All cross-interactions are described via the LJ potential with parameters obtained using Lorentz-Berthelot mixing rules\cite{Allen1990}. Since the aim of the MD simulations is to check and understand whether 1 or 4LG can minimize friction and surface disorder or wear of HDM, the duration is restricted up to when the HDM experiences significant surface disorder and large friction:$\sim$350ps.

Figs.\ref{fig:fig8}a-e,\ref{fig:fig9}a-c compare the computed COF and surface disorder of Co|Pt, 1LG and 4LG. The surface disorder represents the friction induced damage/wear of the HDM, and its value is derived considering 0\% surface disorder before friction measurements. The amount of disorder in HDM is quantified using the centrosymmetry parameter, which is a measure of the local lattice disorder around an atom\cite{Kelcher1998}. This is 0 for a perfect lattice\cite{Kelcher1998}, whereas when point defects exists, i.e. when the symmetry is broken, it assumes a larger positive value\cite{Kelcher1998}.

The simulations indicate that BM develops substrate disorder up to$\sim$14.5\% within 350ps, Figs.\ref{fig:fig8}b),e),\ref{fig:fig9}a). High wear is also observed for BM in the experiments in Fig.\ref{fig:fig3}a). The COF averaged over the final 50ps is$\sim$0.9$\pm$0.1. 1LG reduces the average simulated COF to$\sim$0.2. Fig.\ref{fig:fig8}c shows that 1LG maintains its structural integrity with fraction of disorder$\sim$1\% on average, as a consequence of the low COF between the blocks, indicating higher resistance to wear as compared to BM. 4LG further improves the tribological behaviour. The COF drops to$\sim$0.1 and substrate disorder is reduced to$\sim$0.1\%, much lower than BM, suggesting that the BM surface remains mostly unaffected. Simulations suggest partial transfer of graphene patches to the sapphire ball. The Raman analysis of wear tracks in Figs.\ref{fig:fig5-1},\ref{fig:fig5-2} shows tribo-induced disorder and transfer of carbon to the ball. Thus, the presence of patches containing debris on both surfaces is responsible for the lower COF in 1-4LG-coated HDM, maintaining higher wear resistance than BM.
\section{Conclusions}
1-4LG-coated media can overcome the tribological and corrosion issues of current Co-alloy-based HDM, with laser irradiation stability on FePt-based HDM for future HAMR. The overall performance exceeds that of thicker commercial COC, as well as other amorphous carbons of comparable/higher thicknesses prepared by FCVA and sputtering. Given the tribological, corrosion and thermal stability characteristics coupled with a thickness$\sim$7 times (for 1LG) to$\sim$2 times (for 4LG) lower than state-of-the-art COCs, 1-4LG-based overcoats can meet the requirements for$\sim$4Tb/in$^2$ AD HDDs ($\sim$1.5-1.8nm)\cite{10Marchon2013} and enable the development of ultrahigh AD$\sim$10Tb/in$^2$ and HDDs for HAMR when coupled with BPM. Our results imply that$>$2LG-based coatings could be used as tribological interface for various other materials/devices, such as micro- and nano-electromechanical systems.
\section{Acknowledgments}
We acknowledge funding from the National Research Foundation, Prime Minister's Office, Singapore under its Competitive Research Programme (CRP Award No. NRF-CRP 4-2008-06), the EU Graphene Flagship, EU grant CareRAMM, ERC Grant Hetero2D, EPSRC Grants EP/K01711X/1, EP/K017144/1, EP/N010345/1 and EP/L016057/1, EU grant Neurofibres, the National Energy Research Scientific Computing Center, a DOE Office of Science User Facility, supported by the Office of Science of the U.S. Department of Energy under Contract DE-AC02-05CH11231, the Center for Nanoscale Materials by the U.S. Department of Energy, Office of Science, Office of Basic Energy Sciences, under Contract DE-AC02-06CH11357.

\end{document}